\documentclass{article}

\PassOptionsToPackage{numbers}{natbib}

\usepackage[final]{neurips_2021}

\usepackage[utf8]{inputenc} 
\usepackage[T1]{fontenc}    
\usepackage{hyperref}       
\usepackage{url}            
\usepackage{booktabs}       
\usepackage{amsfonts}       
\usepackage{nicefrac}       
\usepackage{microtype}      

\usepackage[table]{xcolor}
\usepackage{multirow}
\usepackage{chngcntr}
\usepackage{graphicx}
\usepackage{caption}
\usepackage[labelformat=simple]{subcaption}

\usepackage{cleveref}
\crefformat{footnote}{#2\footnotemark[#1]#3}
\newcommand{\ttf}[1]{\textsuperscript{\texttt{#1}}}

\title{DeepRNG: Towards Deep Reinforcement Learning-Assisted Generative Testing of Software}

\author{
Chuan-Yung Tsai\textsuperscript{$\dagger$},
Graham W.~Taylor\textsuperscript{$\dagger\ddagger$}\\
\textsuperscript{$\dagger$}Vector Institute for Artificial Intelligence, Toronto, Canada\\
\textsuperscript{$\ddagger$}School of Engineering, University of Guelph, Guelph, Canada\\
\texttt{kenyon.tsai@vectorinstitute.ai},
\texttt{gwtaylor@uoguelph.ca}
}

\begin{document}

\maketitle

\begin{abstract}
Although machine learning (ML) has been successful in automating various software engineering needs, software testing still remains a highly challenging topic.
In this paper, we aim to improve the generative testing of software by directly augmenting
the random number generator (RNG) with a deep reinforcement learning (RL) agent using an efficient, automatically extractable state representation of the software under test.
Using the Cosmos SDK as the testbed, we show that the proposed DeepRNG framework provides a statistically significant improvement to the testing of the highly complex software library with over 350,000 lines of code.
The source code of the DeepRNG framework is publicly available online.\footnote{\url{https://github.com/cytsai/cosmos-sdk/tree/helios-v1/simapp/rand}\label{footnote:rng}}\textsuperscript{,}\footnote{\url{https://github.com/cytsai/cosmos-sdk-gym}\label{footnote:gym}}
\end{abstract}

\section{Introduction}
Machine learning (ML) has become a popular tool in assisting various software engineering needs \cite{allamanis2018survey}, with successes like code writing services provided by GitHub Copilot \cite{copilot} and OpenAI Codex \cite{chen2021evaluating}.
Software testing, as an equally critical and challenging topic, on the other hand, has yet to receive as much attention in the ML community, even though the number of and the damage caused by software vulnerabilities have been rapidly increasing over the recent years \cite{cve, krasner2021cost}.

Fuzzing is a common technique that uses random inputs to test software.
Although it has played a crucial role in automating software testing for several decades, traditional fuzzing faces many challenges, such as how to efficiently generate valid and effective test inputs \cite{li2018fuzzing}.
Generative testing (also known as property-based testing) on the other hand is mostly paired with a test generator carefully designed based on the specification of the software.
It is similar to a generation-based fuzzer and can often reach deeper program states more effectively \cite{reddy2020quickly}.
Cosmos SDK \cite{kwon2019cosmos}, one of the most used crucial frameworks for building blockchains,\footnote{More than 100 billion USD in digital assets are managed by blockchains built with the Cosmos SDK \cite{tendermint}.} adopts the generative testing approach.
Its built-in test generator generates a random combination of randomly parameterized operations for each block in a simulated blockchain and regularly checks if all invariants (properties) hold true.

In this project, we use the Cosmos SDK as our main testbed for developing a deep reinforcement learning (RL)-assisted generative testing framework.
Although the SDK itself is highly complex with more than 1,500 source files and over 350,000 lines of code, and testing any large-scale software is fundamentally challenging, we demonstrate that our framework, which directly augments the random number generator (RNG) with an efficient deep RL agent, yields a statistically significant improvement to the current testing paradigm.

\subsection{Related Work}
Test coverage is commonly used to measure the progress of testing software, since higher test coverage suggests lower chance of having undetected bugs.
As such, coverage-guided fuzzing has become a highly popular technique and AFL \cite{afl} and its variations are arguably the most widely used coverage-guided fuzzers.
They use genetic algorithms (GAs) to continuously mutate inputs that reached new states of the software under test, which empirically ensures good test coverage.

However, GA-based fuzzers can be very inefficient given that most GAs can only randomly explore the state space, ignoring any underlying structure that may guide the exploration.
As such, NEUZZ \cite{she2019neuzz} proposed to use a neural network (NN) to smoothly approximate the coverage bitmap (binary vector with each bit indicating whether a branch has been executed), whose gradients are then used to perturb the inputs to increase the approximated coverage.

For small-to-medium-sized software with source code available, GMetaExp \cite{dai2019learning} proposed to represent the program as a graph (\textit{e.g.}~abstract syntax tree) with observed coverage information labelled on each vertex.
A graph neural network (GNN) and a long short-term memory network (LSTM) are used to encode the current and past states of the program for the deep RL (DRL) agent to predict actions that maximize the expected reward representing coverage.

Closest to our work, RLCheck \cite{reddy2020quickly} was the first to use RL for generative testing.
It focused on the validity and diversity of the generated test cases using a classical RL algorithm (tabular Q-learning, which cannot handle large state spaces).
It also adopted a manual approach to record recently taken branches (essentially truncated call stacks) as states of the software and recent actions of the RL agent, which require significant changes to the source code.

\begin{table}
\caption{Summary of ML-assisted software testing approaches. Best viewed in color.}
\label{tab:summary}
\centering
\small
\begin{tabular}{rccccc}
\toprule
&
Core\ttf{a} &
\textcolor{blue}{State\ttf{b}} (History\ttf{c}) &
\textcolor{red}{Action\ttf{d}} &
\textcolor{green}{Reward\ttf{e}} &
Automatic\ttf{f} \\
\midrule
AFL \cite{afl} & GA & None & & Coverage & Yes \\
NEUZZ \cite{she2019neuzz} & NN & Coverage Bitmap & Test Input\ttf{g} & Approx.~Coverage & Yes \\
GMetaExp \cite{dai2019learning} & DRL & Coverage Bitmap (\textcolor{blue}{S}) & & Coverage & Unknown \\
\midrule
RLCheck \cite{reddy2020quickly} & RL & Call Stack (\textcolor{red}{A}) & \multirow{2}{*}{Generator Input\ttf{h}} & Trace Diversity & No \\
DeepRNG & DRL & Call Stack (\textcolor{blue}{S}\textcolor{red}{A}\textcolor{green}{R}) & & Coverage & Yes \\
\bottomrule
\end{tabular}\\[1ex]
\footnotesize
\ttf{a}Core learning algorithm.
\ttf{b}Or observation, algorithm's input extracted from the software.
\ttf{c}Past information also used by the algorithm.
\ttf{d}Or mutation/perturbation target, algorithm's output.
\ttf{e}Or objective, optimization target of the algorithm.
\ttf{f}If the software can be automatically instrumented (\textit{e.g.}~to output its states) for the test.
\ttf{g}Input used directly to test the software.
\ttf{h}Input (\textit{e.g.}~random numbers) used by the test generator to test the software.
\end{table}

\subsection{Our Contribution}
We summarize the key features of the related work and our framework in Table~\ref{tab:summary},\footnote{Due to space constraints, we have only listed the three most representative and closely related ML-assisted fuzzers, AFL, NEUZZ and GMetaExp. Please see \cite{saavedra2019review,wang2020systematic} for more examples.} and our main contributions as follows:
\begin{itemize}
\item We propose to use the (full) call stack of the software (see Fig.~\ref{fig:helios} and \ref{fig:bst} for examples) as the state representation which is effective and lightweight (compared to \textit{e.g.}~a coverage bitmap), and implement and open-source the library for automatic instrumentation.\cref{footnote:rng}
\item We extend the state-history representations proposed by the previous work by using a deep RL architecture which can flexibly encode history of past states, actions and rewards without needing to truncate older history (as in \textit{e.g.}~RLCheck \cite{reddy2020quickly}).
\item We show that the DeepRNG framework can achieve better coverage for the Cosmos SDK in a statistically significant way, as well as identifying previously unknown bugs.\footnote{\url{https://github.com/cosmos/cosmos-sdk/discussions/9178\#discussioncomment-763961}\label{footnote:bugs}}
\end{itemize}
\pagebreak

\section{DeepRNG Framework}
\begin{figure}
\centering
\includegraphics[width=0.8\textwidth]{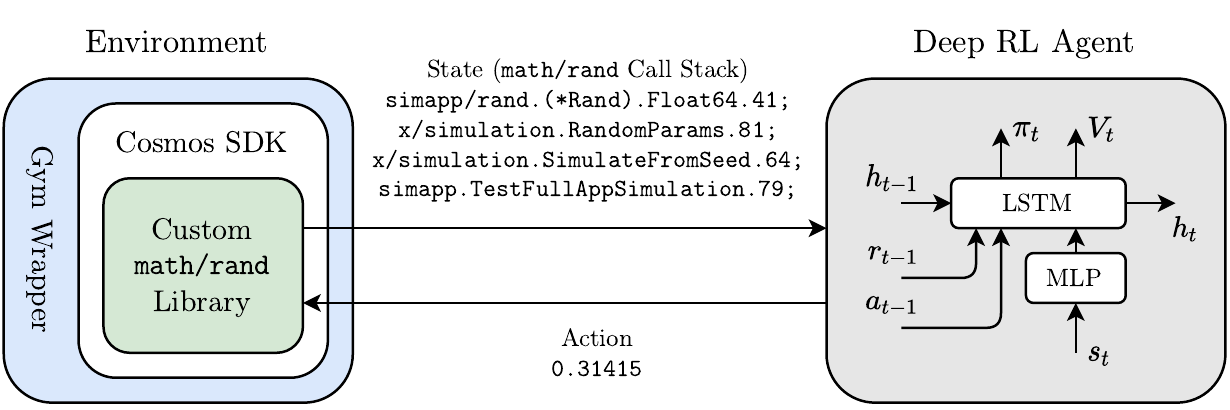}
\caption{
The DeepRNG framework, where the colored modules (the custom RNG library and the OpenAI Gym wrapper) are the main software contributions of this project.
The custom RNG library routes function calls to an external RL agent, with the call stack information passed as the states of the software under test, and the coverage information and exit codes (not shown in the figure) passed as the rewards.
The RL agent then returns random numbers as actions that are sampled from a distribution learned to maximize the expected reward, conditioned on an encoding of the current states and optionally the past states, actions and rewards.
The states, rewards and actions are passed as strings over \texttt{stdout}, \texttt{stdin} and/or named pipes, which are mainly handled by the OpenAI Gym wrapper to ensure the communication starts and ends correctly. 
In this example, the Cosmos SDK's \texttt{TestFullAppSimulation} routine (the ``main function'' for testing) entered the \texttt{SimulateFromSeed} subroutine, which called \texttt{RandomParams} that in turn requested a random \texttt{Float64} (between $0.0$ and $1.0$) from Go's RNG library, \texttt{math/rand}.
The RL agent then returned \texttt{0.31415} as its action.
Best viewed in color.}
\label{fig:helios}
\end{figure}

The proposed DeepRNG framework is illustrated in Fig.~\ref{fig:helios}.
The two new modules designed for this project, the custom RNG library implemented in Go\cref{footnote:rng} and its corresponding OpenAI Gym wrapper,\cref{footnote:gym} are both open-source and publicly available online.
The custom RNG library can also reproduce simulations from log files without launching an RL agent, which is essential for reporting bugs.
To achieve the goal of automatically extracting the call stack as the state of the software, we use the \texttt{runtime.CallersFrames} function in Go.
Similar functions can be found in most programming languages \cite{wiki:Stack_trace}, meaning that the DeepRNG framework is in theory extensible to most languages.

For the deep RL agent, we adopt the IMPALA algorithm \cite{espeholt2018impala} as well as its neural network architecture which flexibly supports encoding of the current state ($s_t$) and optionally the history of the past states, actions and rewards ($h_{t-1}$, $a_{t-1}$, $r_{t-1}$) of the software and the agent, using a multilayer perceptron (MLP) and a long short-term memory network (LSTM).
In addition, the IMPALA agent is highly computationally efficient by design and provides the fastest training speed among all the deep RL agents we have tested.

\section{Experiments}
We use the IMPALA agent implemented in RLlib \cite{liang2018rllib} for all of our experiments on a workstation with an Intel Core i7-10700 CPU, 64 GB of main memory and an Nvidia GeForce RTX 3090 GPU.\footnote{We use 8 parallel environments and 8 CPU-based actors to collect trajectories for one GPU-based learner.}
We focus on comparing different state-history representations to validate our approach and to showcase our framework's ability to better test real-world software (in this paper, the Cosmos SDK).

\begin{figure}
\begin{subfigure}[t]{0.49\textwidth}
\includegraphics[width=\textwidth]{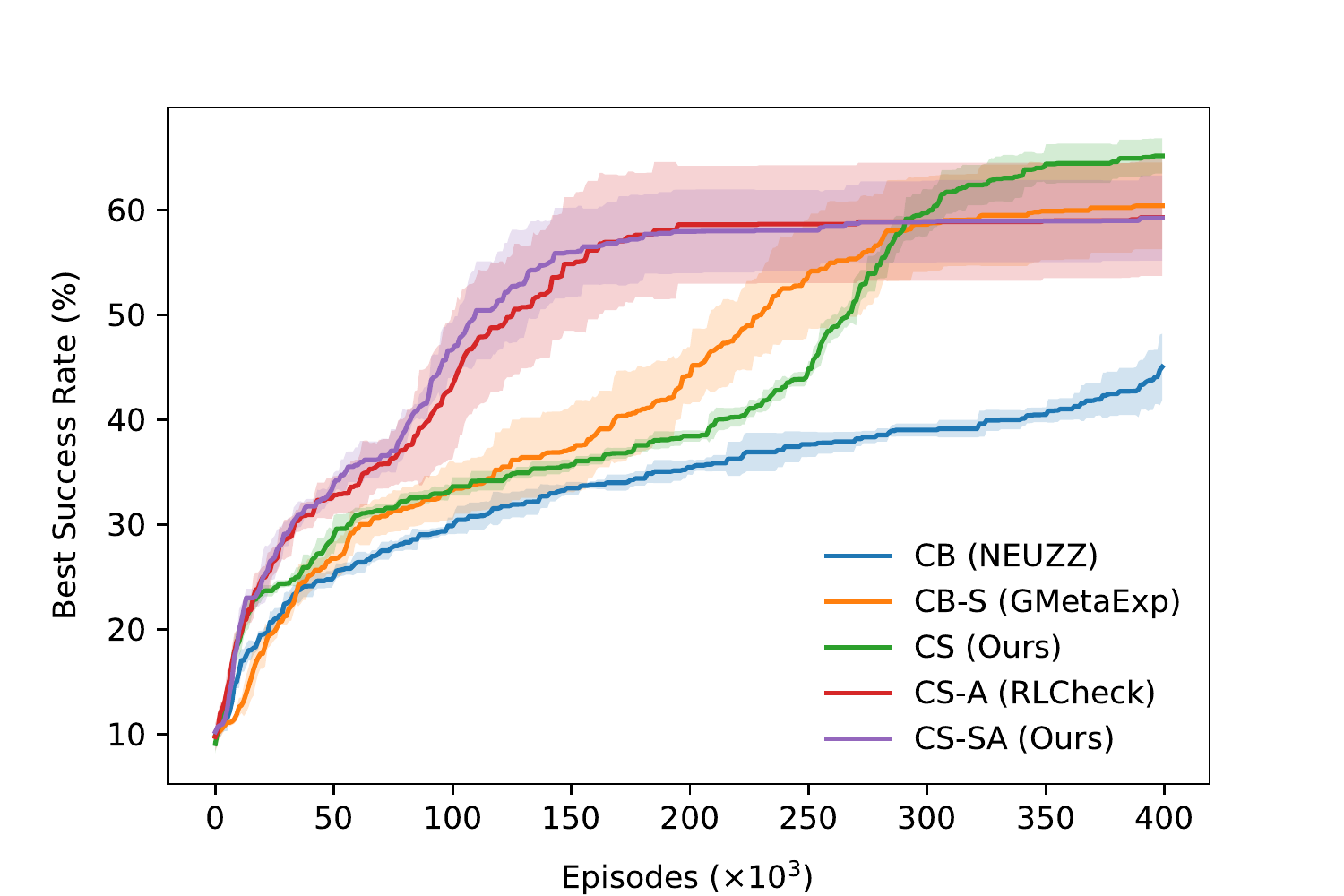}
\caption{\centering Success rate (\textit{i.e.}~average reward) of generating valid binary search trees (BSTs).}
\label{fig:rate}
\end{subfigure}
\hfill
\begin{subfigure}[t]{0.49\textwidth}
\includegraphics[width=\textwidth]{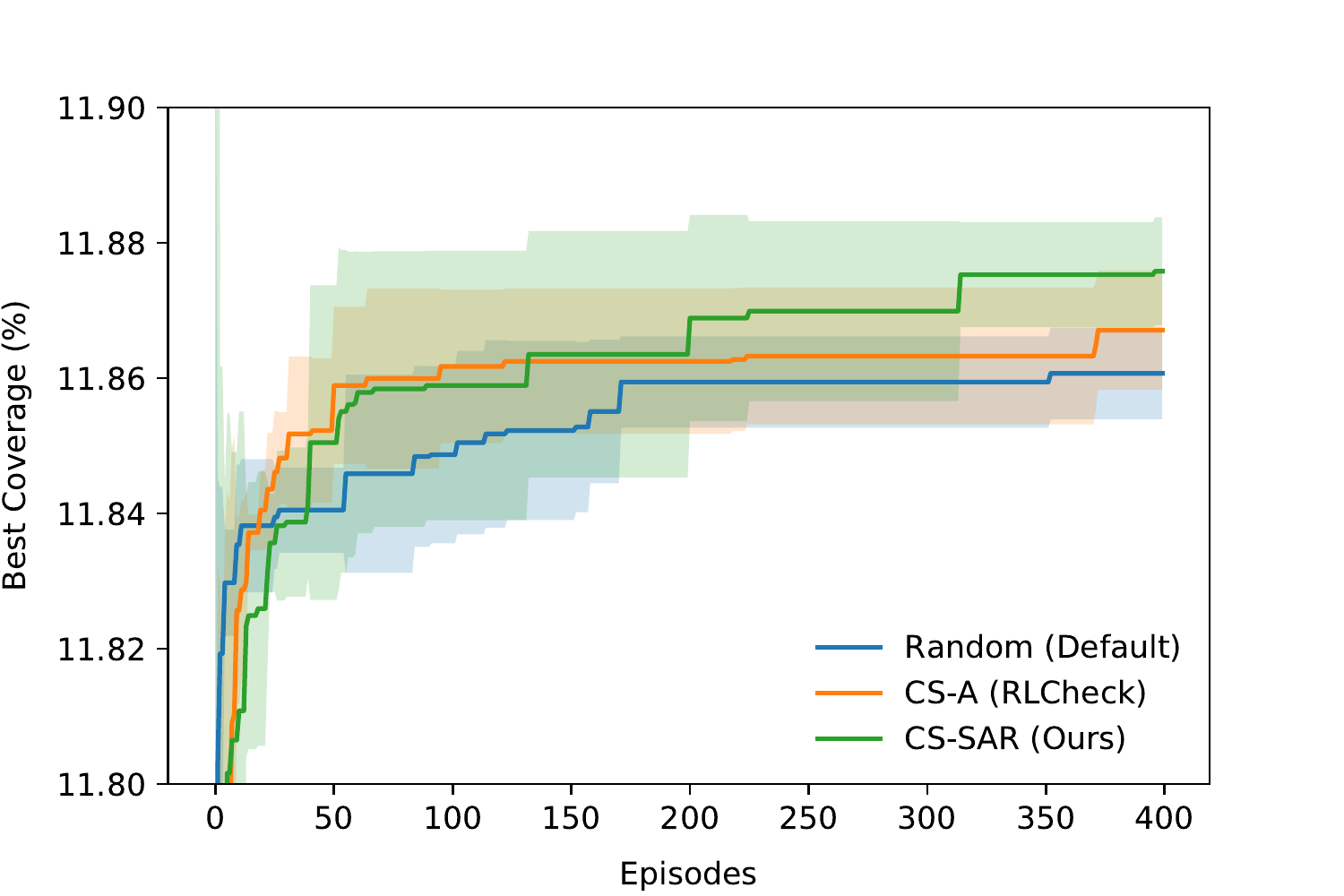}
\caption{\centering Test coverage (\textit{i.e.}~reward) of the Cosmos SDK.}
\label{fig:coverage}
\end{subfigure}
\caption{
Experimental results.
Both subfigures plot the best value achieved at a given episode since the beginning (also known as the maximum reward).
For each curve (one state-history representation), state can be coverage bitmap (CB) or call stack (CS), and history can be any combination of past states (S), actions (A) and rewards (R).
Best viewed in color.
}
\end{figure}

\subsection{Binary Search Tree Generation}
Inspired by the binary search tree (BST) test case generation problem in \cite{reddy2020quickly}, we create a similar toy problem focusing on randomly generating valid BSTs,\footnote{A valid BST is a binary tree whose internal nodes each store a number greater than all the numbers in the node's left subtree and less than those in its right subtree.} where randomly numbered nodes are recursively generated until \texttt{TREE\_DEPTH} and each node’s left and right subtrees are independently pruned with a probability of \texttt{TREE\_PRUNE}.
We set $\texttt{TREE\_DEPTH} = 4$, $\texttt{TREE\_PRUNE} = 0.5$ and the agent's action space to integers between $1$ and $31$ such that the optimal solution (agent always able to generate valid BSTs) is achievable.
Worth to note, even though the optimal solution is obvious to experienced programmers, the RL agent can only discover the solution using the very sparse binary rewards ($0.0$ for invalid BSTs, $1.0$ for valid BSTs) that it receives at the end of each run (episode).
More details about this problem and its optimal solution can be found in the appendix.

We train the IMPALA agent using different state-history representations for 0.4 million episodes (around 2 million environment steps) and 5 repeated runs per representation.
The results are shown in Fig.~\ref{fig:rate}.
Here, the coverage bitmap (CB) representation consists of a 31-bit binary vector indicating the coverage status of all 31 nodes within the tree, and the call stack (CS) representation consists of a 31-bit one-hot vector indicating which one of the 31 nodes is currently being visited.
Interestingly, although the CB representation carries more information (than the CS), it actually performs worse (or no better).\footnote{We have also tested combining both representations, which yields results similar to using the CS alone.}
Moreover, among all CS-based representations, not using any history actually performs the best.
This suggests that more information is not necessarily better (perhaps particularly so for RL problems where credit assignment is fundamentally challenging) and having a flexible framework based on efficient representations to facilitate the exploration of configurations is more critical.

\subsection{Cosmos SDK}
\begin{table}
\caption{Cosmos SDK test coverage results.}
\label{tab:results}
\centering
\small
\begin{tabular}{lcc}
\toprule
& Best Coverage ($\%$) & $p$-value \\
\midrule
Random (Default) & $11.861\pm0.007$ & Baseline \\
\midrule
CS             & $11.864\pm0.009$ & $0.584$ \\
CS-S           & $11.870\pm0.011$ & $0.139$ \\
\rowcolor{yellow}
CS-SR          & $11.883\pm0.018$ & $0.032$ \\
CS-A (RLCheck) & $11.867\pm0.009$ & $0.234$ \\
CS-AR          & $11.869\pm0.007$ & $0.088$ \\
\rowcolor{yellow}
CS-SA          & $11.876\pm0.010$ & $0.024$ \\
\rowcolor{yellow}
CS-SAR         & $11.876\pm0.008$ & $0.012$ \\
\bottomrule
\end{tabular}
\end{table}

For the Cosmos SDK experiments, we train the IMPALA agent using different state-history representations for 400 episodes (around 10 million environment steps) and 5 repeated runs per representation.
Because the coverage bitmap of the Cosmos SDK (like any other large-scale software) is too large (around 78,000-bits long) to be processed efficiently, we only use the call stack (CS) as the state representation, which is simply encoded as a 158-bit one-hot vector indicating which one of the 158 unique \texttt{math/rand} call stacks is currently being observed.
The agent is tasked to generate actions (random numbers) within $[0,1)$, which can be post-processed to meet different types of RNG requests (\textit{e.g.}~scaled and rounded for random integers within different ranges).
During each episode, the agent is continuously rewarded with the increases in coverage\footnote{Per-step reward is the difference between previous-step and current-step coverages. Coverage is measured using the \texttt{testing.Coverage} function in Go, which estimates the branch coverage \cite{cover}.} (which sum to the final coverage between $0.0$ and $1.0$) and a final reward of $1.0$ if the simulation fails (\textit{i.e.}~a bug was encountered).

The results are shown in Fig.~\ref{fig:coverage} and summarized in Table~\ref{tab:results}.
Compared to the default random testing of the Cosmos SDK, using the call stack (CS) representation combined with the state, action and reward (SAR) history achieves the most statistically significant improvement based on the \textit{t}-test (along with two other also statistically significant configurations as highlighted in Table~\ref{tab:results}).
Incorporating the reward history could allow the agent to ``know'' how much of the software has (not) been covered and to adapt its strategy, which does seem to improve or robustify the coverage results.
In addition to improving the coverage, the DeepRNG framework also has identified at least two types of bugs (invalid initialization and unexpected timeout) in the Cosmos SDK.\cref{footnote:bugs}

\section{Discussion}
In this paper, we propose the DeepRNG framework and experimentally validate its effectiveness by showing promising results for testing the Cosmos SDK.
Below are some of our future directions.
\begin{itemize}
\item For the Cosmos SDK, although the improvement in test coverage is statistically significant, the numerical increase is actually quite small.
This is mainly due to the fact that the SDK has many modules not reachable by its current test generator.
Filtering out unreachable (or unimportant) modules may simplify the problem and lead to better results.
\item To further optimize the test coverage, it's also possible to completely disregard the ``exploitation phase'' of the RL algorithm, since exploring new states is the only source of rewards.
The exploration phase of \citep{ecoffet2019go, ecoffet2021first} may be used to this end.
\end{itemize}

\section*{Acknowledgments}
We thank the Interchain Foundation for funding this project, and Ethan Buchman and members of Informal Systems for valuable suggestions.
Resources used in preparing this research were provided, in part, by the Province of Ontario, the Government of Canada through CIFAR, and companies sponsoring the Vector Institute.

{\small
\bibliographystyle{plainnat}
\setlength{\bibsep}{1.2ex}
\bibliography{references}

\begin{thebibliography}{20}
\providecommand{\natexlab}[1]{#1}
\providecommand{\url}[1]{\texttt{#1}}
\expandafter\ifx\csname urlstyle\endcsname\relax
  \providecommand{\doi}[1]{doi: #1}\else
  \providecommand{\doi}{doi: \begingroup \urlstyle{rm}\Url}\fi

\bibitem[Allamanis et~al.(2018)Allamanis, Barr, Devanbu, and
  Sutton]{allamanis2018survey}
Miltiadis Allamanis, Earl~T Barr, Premkumar Devanbu, and Charles Sutton.
\newblock A survey of machine learning for big code and naturalness.
\newblock \emph{ACM Computing Surveys (CSUR)}, 51\penalty0 (4):\penalty0 1--37,
  2018.

\bibitem[Chen et~al.(2021)Chen, Tworek, Jun, Yuan, Ponde, Kaplan, Edwards,
  Burda, Joseph, Brockman, et~al.]{chen2021evaluating}
Mark Chen, Jerry Tworek, Heewoo Jun, Qiming Yuan, Henrique Ponde, Jared Kaplan,
  Harri Edwards, Yura Burda, Nicholas Joseph, Greg Brockman, et~al.
\newblock Evaluating large language models trained on code.
\newblock \emph{arXiv preprint arXiv:2107.03374}, 2021.

\bibitem[Corporation(2021)]{cve}
MITRE Corporation.
\newblock {CVE} (common vulnerabilities and exposures) details.
\newblock \url{https://www.cvedetails.com/browse-by-date.php}, 2021.

\bibitem[Dai et~al.(2019)Dai, Li, Wang, Singh, Huang, and
  Kohli]{dai2019learning}
Hanjun Dai, Yujia Li, Chenglong Wang, Rishabh Singh, Po-Sen Huang, and Pushmeet
  Kohli.
\newblock Learning transferable graph exploration.
\newblock \emph{Advances in Neural Information Processing Systems},
  32:\penalty0 2518--2529, 2019.

\bibitem[Ecoffet et~al.(2019)Ecoffet, Huizinga, Lehman, Stanley, and
  Clune]{ecoffet2019go}
Adrien Ecoffet, Joost Huizinga, Joel Lehman, Kenneth~O Stanley, and Jeff Clune.
\newblock Go-explore: a new approach for hard-exploration problems.
\newblock \emph{arXiv preprint arXiv:1901.10995}, 2019.

\bibitem[Ecoffet et~al.(2021)Ecoffet, Huizinga, Lehman, Stanley, and
  Clune]{ecoffet2021first}
Adrien Ecoffet, Joost Huizinga, Joel Lehman, Kenneth~O Stanley, and Jeff Clune.
\newblock First return, then explore.
\newblock \emph{Nature}, 590\penalty0 (7847):\penalty0 580--586, 2021.

\bibitem[Espeholt et~al.(2018)Espeholt, Soyer, Munos, Simonyan, Mnih, Ward,
  Doron, Firoiu, Harley, Dunning, et~al.]{espeholt2018impala}
Lasse Espeholt, Hubert Soyer, Remi Munos, Karen Simonyan, Vlad Mnih, Tom Ward,
  Yotam Doron, Vlad Firoiu, Tim Harley, Iain Dunning, et~al.
\newblock {IMPALA}: Scalable distributed deep-{RL} with importance weighted
  actor-learner architectures.
\newblock In \emph{International Conference on Machine Learning}, pages
  1407--1416. PMLR, 2018.

\bibitem[GitHub(2021)]{copilot}
GitHub.
\newblock {GitHub Copilot}.
\newblock \url{https://copilot.github.com}, 2021.

\bibitem[Inc(2021)]{tendermint}
Tendermint Inc.
\newblock Tendermint’s proposal for a new versioning of {Cosmos SDK}.
\newblock
  \url{https://medium.com/tendermint/tendermints-proposal-for-a-new-versioning-of-cosmos-sdk-d52a01976852},
  2021.

\bibitem[Krasner(2021)]{krasner2021cost}
Herb Krasner.
\newblock The cost of poor software quality in the {US}: A 2020 report.
\newblock In \emph{Proc. Consortium Inf. Softw. QualityTM (CISQTM)}, 2021.

\bibitem[Kwon and Buchman(2019)]{kwon2019cosmos}
Jae Kwon and Ethan Buchman.
\newblock Cosmos whitepaper.
\newblock \url{https://cosmos.network/resources/whitepaper}, 2019.

\bibitem[Li et~al.(2018)Li, Zhao, and Zhang]{li2018fuzzing}
Jun Li, Bodong Zhao, and Chao Zhang.
\newblock Fuzzing: a survey.
\newblock \emph{Cybersecurity}, 1\penalty0 (1):\penalty0 1--13, 2018.

\bibitem[Liang et~al.(2018)Liang, Liaw, Nishihara, Moritz, Fox, Goldberg,
  Gonzalez, Jordan, and Stoica]{liang2018rllib}
Eric Liang, Richard Liaw, Robert Nishihara, Philipp Moritz, Roy Fox, Ken
  Goldberg, Joseph Gonzalez, Michael Jordan, and Ion Stoica.
\newblock {RLlib}: Abstractions for distributed reinforcement learning.
\newblock In \emph{International Conference on Machine Learning}, pages
  3053--3062. PMLR, 2018.

\bibitem[Pike(2013)]{cover}
Rob Pike.
\newblock The {Go} blog: The cover story.
\newblock \url{https://go.dev/blog/cover}, 2013.

\bibitem[Reddy et~al.(2020)Reddy, Lemieux, Padhye, and Sen]{reddy2020quickly}
Sameer Reddy, Caroline Lemieux, Rohan Padhye, and Koushik Sen.
\newblock Quickly generating diverse valid test inputs with reinforcement
  learning.
\newblock In \emph{2020 IEEE/ACM 42nd International Conference on Software
  Engineering (ICSE)}, pages 1410--1421. IEEE, 2020.

\bibitem[Saavedra et~al.(2019)Saavedra, Rodhouse, Dunlavy, and
  Kegelmeyer]{saavedra2019review}
Gary~J Saavedra, Kathryn~N Rodhouse, Daniel~M Dunlavy, and Philip~W Kegelmeyer.
\newblock A review of machine learning applications in fuzzing.
\newblock \emph{arXiv preprint arXiv:1906.11133}, 2019.

\bibitem[She et~al.(2019)She, Pei, Epstein, Yang, Ray, and Jana]{she2019neuzz}
Dongdong She, Kexin Pei, Dave Epstein, Junfeng Yang, Baishakhi Ray, and Suman
  Jana.
\newblock {NEUZZ}: Efficient fuzzing with neural program smoothing.
\newblock In \emph{2019 IEEE Symposium on Security and Privacy (SP)}, pages
  803--817. IEEE, 2019.

\bibitem[Wang et~al.(2020)Wang, Jia, Liu, Huang, and Liu]{wang2020systematic}
Yan Wang, Peng Jia, Luping Liu, Cheng Huang, and Zhonglin Liu.
\newblock A systematic review of fuzzing based on machine learning techniques.
\newblock \emph{PloS one}, 15\penalty0 (8):\penalty0 e0237749, 2020.

\bibitem[Wikipedia(2021)]{wiki:Stack_trace}
Wikipedia.
\newblock Stack trace.
\newblock \url{https://en.wikipedia.org/wiki/Stack_trace}, 2021.

\bibitem[Zalewski(2013)]{afl}
Michał Zalewski.
\newblock american fuzzy lop.
\newblock \url{https://lcamtuf.coredump.cx/afl}, 2013.

\end{thebibliography}
}

\appendix
\counterwithin{figure}{section}
\section{Appendix}

\begin{figure}[h]
\centering
\includegraphics[width=0.95\textwidth]{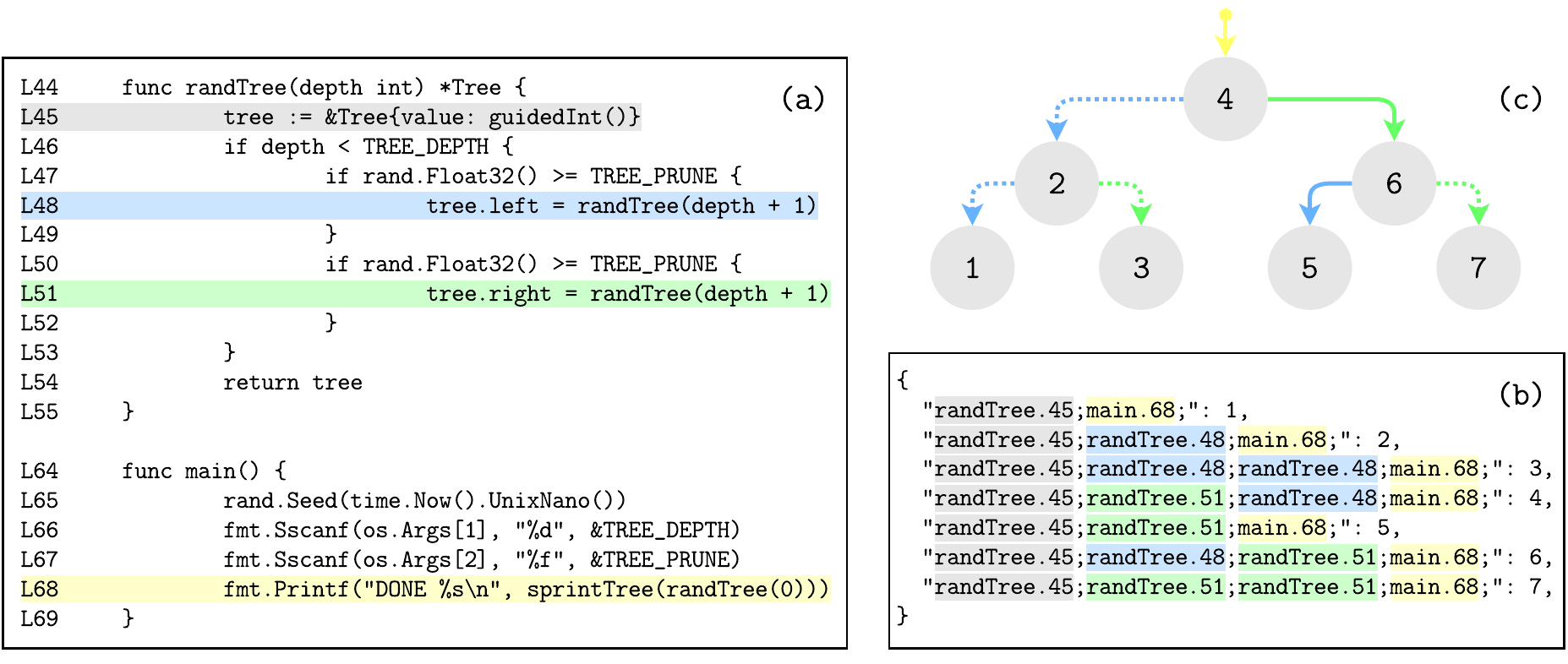}
\caption{
Binary search tree (BST) generation.
(a) Go program of a random BST generator where \texttt{randTree} is called recursively to generate randomly numbered nodes until \texttt{TREE\_DEPTH} and each node's left and right subtrees are pruned with a probability of \texttt{TREE\_PRUNE} independently.
Following the DeepRNG framework, the random number for each node is generated using a custom \texttt{guidedInt} function, which passes its state (call stack) to an external RL agent each time being called.
The agent is then trained to correspond with a random number that maximizes its final reward, the validity of the BST.
(b) All 7 possible states (call stacks) for $\texttt{TREE\_DEPTH} = 2$.
(c) The optimal solution of an agent with an action space of integers between $1$ and $7$.
\textit{I.e.}~an agent that learns the state-action map $f:s{\to}a$ $\forall(s,a) \in \lbrace(1,4),(2,2),(3,1),(4,3),(5,6),(6,5),(7,7)\rbrace$ always gets a reward (a valid BST, like $4\rightarrow(5\leftarrow6)$ in the figure) regardless of the tree shape and order of new nodes (due to random pruning it cannot control).
}
\label{fig:bst}
\end{figure}

\end{document}